\documentclass[]{piparticle-final}
\usepackage{graphicx}
\usepackage{amsmath}
\usepackage{epstopdf}

\usepackage{cite} % To compress citation ranges
\begin{document}

\volume{5}               % To be inserted by Editor
\articlenumber{050004}   % To be inserted by Editor
\journalyear{2013}       % To be inserted by Editor
\editor{G. Mindlin}   % To be inserted by Editor
%\reviewers{Reviewer's name}  % To be inserted by Editor
\received{2 March 2013}     % To be inserted by Editor
\accepted{5 June 2013}   % To be inserted by Editor
\runningauthor{M. F. Assaneo \itshape{et al.}}  % To be inserted by Editor
\doi{050004}         % To be inserted by Editor

\title{Revisiting the two-mass model of the vocal folds}

% Institution references with \cite are inserted after \maketitle in theaffiliation enviroment
\author{M. F. Assaneo,\cite{inst1}\thanks{E-mail: florencia@df.uba.ar} \hspace{0.5em}  
        M. A. Trevisan\cite{inst1}\thanks{E-mail: marcos@df.uba.ar}}

\pipabstract{
Realistic mathematical modeling of voice production has been recently boosted by applications to different fields like 
bioprosthetics, quality speech synthesis and pathological diagnosis. In this work, we revisit a two-mass model of the 
vocal folds that includes accurate fluid mechanics for the air passage through the folds and nonlinear properties of the 
tissue. 
We present the bifurcation diagram for such a system, focusing  on the dynamical 
properties of two regimes of interest: the onset of oscillations and the normal phonation regime. We also show theoretical 
support to the nonlinear nature of the elastic properties of the folds tissue by comparing theoretical isofrequency curves 
with reported experimental data.}

\maketitle

\blfootnote{
\begin{theaffiliation}{99}
   \institution{inst1} Laboratorio de Sistemas Din\'amicos, Depto. de F\'isica, FCEN, Universidad de Buenos Aires. Pabell\'on I, Ciudad Universitaria, 1428EGA Buenos Aires, Argentina.
\end{theaffiliation}
}

\section{Introduction}
\label{intro}

In the last decades, a lot of effort was devoted to develop a mathematical model for voice production. 
The first steps were made by Ishizaka and Flanagan \cite{1}, approximating each vocal fold by two coupled 
oscillators, which  provide the basis of the well known two-mass model. 
This simple model reproduces many essential features of the voice production, like the onset of self 
sustained oscillation of the folds and the shape of the glottal pulses. 

Early analytical treatments were restricted to small amplitude oscillations, allowing a dimensional reduction 
of the problem. In particular, a two dimensional approximation known as the flapping model was widely adopted 
by the scientific community, based on the assumption of a transversal wave propagating along the vocal folds  
\cite{6,7}. Moreover, this model was also used to successfully explain most of the features present in birdsong 
\cite{2,3}.  

Faithful modeling of the vocal folds has recently found new challenges: realistic articulatory speech synthesis 
\cite{11,12,17}, diagnosis of pathological behavior of the folds \cite{13,14} and bioprosthetic applications 
\cite{15}.  Within this framework, the 4-dimensional two-mass model was revisited 
and modified. Two main improvements are worth noting: a realistic description of the vocal fold collision 
\cite{8,9} and an accurate fluid mechanical description of the glottal flow, allowing a proper treatment of 
the hydrodynamical force acting on the folds \cite{10, 17}.

In this work, we revisit the two-mass model developed by Lucero and Koenig \cite{12}. 
This choice represents a good compromise between mathematical simplicity and diversity of physical phenomena
acting on the vocal folds, including the main mechanical and fluid effects that are partially found in other 
models \cite{10, 8}. It was also successfully used to reproduce experimental temporal patterns of glottal airflow. 
Here, we extend the analytical study of this system: we present a bifurcation diagram, explore the dynamical 
aspects of the  oscillations at the onset and normal phonation and  study the isofrequency curves of the model.

This work is organized as follows: in the second section, we describe the model. In the third section, 
we present the bifurcation diagram, compare our solutions with those of the flapping model approximation and
analyze the isofrecuency curves. In the fourth and last section, we  discuss our results.

\section{The model}
\label{model}

Each vocal fold is modeled as two coupled damped oscillators, as sketched in Fig. 1.

\begin{figure}[!ht]
\begin{center}
\includegraphics[width=0.47\textwidth]{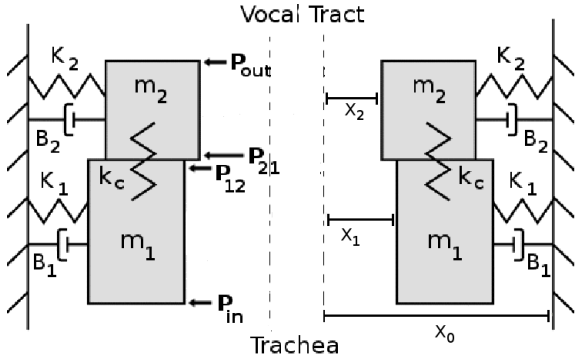}
\end{center}
\caption{Sketch of the two-mass model of the vocal folds. Each fold is represented by masses $m_1$ and $m_2$ coupled to
each other by a restitution force $k_c$ and to the laryngeal walls by $K_1$ and $K_2$ (and dampings $B_1$ and $B_2$), 
respectively. The displacement of each mass from the resting position $x_0$ is represented by $x_1$ and $x_2$.
The different aerodynamic pressures $P$ acting on the folds are described in the text.} \label{figure1}
\end{figure}

Assuming symmetry with respect to the saggital plane, the left and right mass systems are identical (Fig. \ref{figure1}) 
and the equation of motion for each mass reads

\begin{align}
\label{motion}
\dot{x_i}&= y_i \\
\dot{y_i}&= \frac{1}{m_i} \left[f_i - K_i(x_i) - B_i(x_i, y_i) - k_c(x_i-x_j)\right], \nonumber 
\end{align} 

\noindent for $i,j=1$ or 2 for lower and upper masses, respectively. $K$ and $B$ represent the restitution and damping of 
the folds tissue, $f$ the hydrodynamic force, $m$ is the mass and $k_c$ the coupling stiffness. The horizontal displacement 
from the rest position $x_0$ is represented by $x$. 

We use a cubic polynomial for the restitution term [Eq. (\ref{restitution})], adapted from \cite{1,12}. The term with a 
derivable \textit{step-like} function $\Theta$ [Eq. (\ref{theta})] accounts for the increase in the stiffness introduced 
by the collision of the folds. The restitution force reads  
 
\begin{align}
\label{restitution}
 K_i&(x_i)=k_i x_i(1+100 {x_i}^2)   \\
 &+\Theta\left(\frac{x_i+x_0}{x_0}\right) 3 k_i (x_i + x_0)[1+500 {(x_i+x_0)}^2],   \nonumber
\end{align}
with

\begin{align}
\label{theta}
\Theta(x) = \left\{
\begin{array}{rl}
0 & \text{if } x \leq 0\\
\frac{x^2}{8\text{ }10^{-4}+x^2} & \text{if } x > 0
\end{array}  \right.,
\end{align}

\noindent where $x_0$ is the rest position of the folds.

For the damping force, we have adapted the expression proposed in \cite{12}, making it derivable, arriving  at the 
following equation:

\begin{align}
B_i(x_i)=& \\
& \left[1 +\Theta\left(\frac{x_i+x_0}{x_0}\right)\frac{1}{\epsilon_i}\right] r_i(1+850{x_i}^2) y_i,  \nonumber
\end{align}

\noindent where $r_i=2\epsilon_i \sqrt{k_i m_i} $, and $\epsilon_i$ is the damping ratio. 

In order to describe the hydrodynamic force that the airflow exerts on the vocal folds, we have adopted the standard 
assumption of small inertia of the glottal air column and the model of the boundary layer developed in \cite{12,15,10}. 
This model assumes a one-dimensional, quasi-steady incompressible airflow from the trachea to a \textit{separation point}. 
At this point, the flow separates from the tissue surface to form a free jet where the turbulence dissipates the airflow 
energy. It has been experimentally shown that the position of this point depends on the glottal profile. As described 
in \cite{10}, the separation point located at the glottal exit shifts down to the boundary between masses $m_1$ and 
$m_2$ when the folds profile becomes more divergent than a threshold [Eq. (\ref{visco_partida})]. 

Viscous losses are modeled according to a bi-dimensional Poiseuille flow [Eqs. (\ref{visco}) and (\ref{visco_partida})]. 
The equations for the pressure inside the glottis are  
      
\begin{align}
\label{nl}
P_{in}&=P_s + \frac{\rho u_g ^2}{2a_1 ^2},  \\
\label{visco}
P_{12}&= P_{in}- \frac {12 \mu u_g d_1 l_g ^2}{a_1 ^3}, \\ 
\label{visco_partida}
P_{21}&= \left\{
\begin{array}{rl}
\frac {12 \mu u_g d_2 l_g ^2}{a_2 ^3} +P_{out} & \text{if } a_2 > k_s a_1 \\
0 & \text{if } a_2 \leq k_s a_1  
\end{array}, \right. \\
\label{p_out}
P_{out}&=0.
\end{align}

As sketched in Fig. \ref{figure1}, the pressures exerted by the airflow are: $P_{in}$ at the entrance of the glottis, 
$P_{12}$ at the upper edge of $m_1$, $P_{21}$ at the lower edge of $m_2$, $P_{out}$ at the entrance of the vocal 
tract and $P_s$ the subglottal pressure. 

The width of the folds (in the plane normal to Fig. \ref{figure1}) is $l_g$; $d_1$ and $d_2$ are the lengths of the 
lower and upper masses, respectively. $a_i$ are the cross-sections of the glottis, $a_i=2l_g(x_i+x_0)$; $\mu$ and 
$\rho$ are the viscosity and density coefficient of the air; $u_g$ is the airflow inside the glottis, and $k_s=1.2$ 
is an experimental coefficient. We also assume no losses at the glottal entrance [Eq. (\ref{nl})], and zero pressure 
at the entrance of the vocal tract [Eq. (\ref{p_out})]. 

The hydrodynamic force acting on each mass reads: 

\begin{align}
\label{f1}
f_1 = \left\{
\begin{array}{rl}
d_1 l_g P_s & \text{if } x_1 \leq -x_0 \text{ or } x_2 \leq -x_0 \\ 
\frac {P_{in}+P_{12}}{2}  & \text{in other case} \
\end{array} \right.
\end{align}

\begin{align}
\label{f2}
f_2 = \left\{
\begin{array}{rl}
d_2 l_g P_s & \text{if } x_1 > -x_0 \text{ and } x_2 \leq -x_0 \\ 
0 & \text{if } x_1 \leq -x_0 \\
\frac {P_{21}+P_{out}}{2}  & \text{in other case} \
\end{array} \right. 
\end{align}

Following \cite{1,12,14}, these functions represent opening, partial closure and total closure of the glottis. 
Throughout this work, piecewise functions $P_{21}$, $f_1$ and $f_2$ are modeled using the derivable 
step-like function $\Theta$ defined in Eq. (\ref{theta}).

\section{Analysis of the model}
\label{analysis}

\subsection{Bifurcation diagram}
\label{bifurcation}

The main anatomical parameters that can be actively controlled during the vocalizations are the subglottal pressure 
$P_s$ and the folds tension controlled by the laryngeal muscles. In particular, the action of the thyroarytenoid 
and the cricothyroid muscles control the thickness and the stiffness of folds. Following \cite{1}, this effect is 
modeled by a parameter $Q$ that scales the mechanic properties of the folds by a cord-tension parameter: 
$k_c=Q k_{c0}$, $k_i=Q k_{i0}$ and $m_i=\frac {m_{i0}}{Q}$. 
We therefore performed a bifurcation diagram using these two standard control parameters $P_s$ and $Q$. 

\begin{figure}[!ht]
\begin{center}
\includegraphics[width=0.46\textwidth]{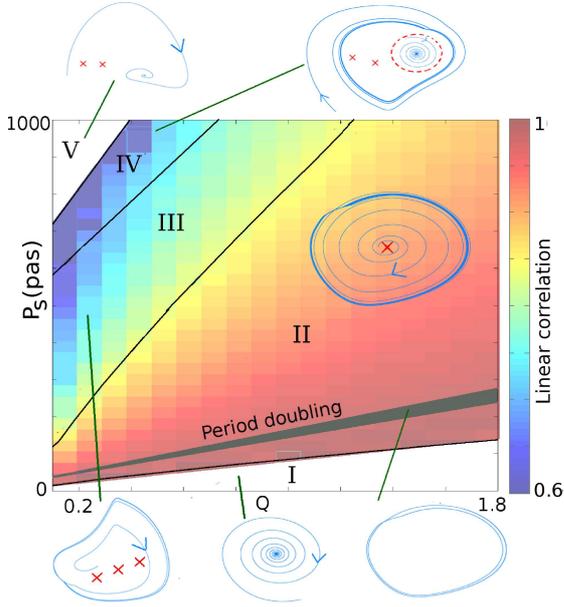}
\end{center}
\caption{Bifurcation diagram in the plane of subglottal pressure and fold tension ($Q$,$P_s$). The insets are 
two-dimensional projections of the flow on the ($v_1$,$x_1$) plane, the red crosses represent unstable fixed 
points and the dotted lines unstable limit cycles. Normal voice occurs at $(Q,P_s)\sim(1,800)$. The color code 
represents the linear correlation between $(x_1-x_2)$ and $(y_1+y_2)$: from dark red for $R=1$ to
dark blue for $R=0.6$. This diagram was developed with the help of AUTO continuation software \cite{20}. 
The rest of the parameters were fixed at $m_1=0.125$ g, $m_2=0.025$ g, $k_{10}= 80$ N/m, $k_{20}= 8$ N/m, $k_c= 25$ N/m, 
$\epsilon_1 = 0.1$, $\epsilon_2= 0.6$, $l_g= 1.4$ cm, $d_1= 0.25$ cm, $d_2=0.05$ cm and $x_0= 0.02$ cm.}
\label{figure2}
\end{figure}

\begin{figure}[!ht]
\begin{center}
\includegraphics[width=0.45\textwidth]{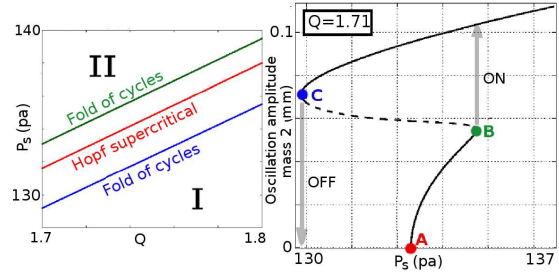}
\end{center}
\caption{Hysteresis at the oscillation onset-offset. Left panel: zoom of the interface between regions I and II.
The blue and green lines represent folds of cycles (saddle-node bifurcations in the map). 
The red line is a supercritical Hopf bifurcation. Right panel: the oscillation amplitude of $x_2$ as a function of 
the subglottal pressure $P_s$, at $Q=1.71$. The continuation of periodic solutions was realized with the 
AUTO software package \cite{20}.}
\label{figure3}
\end{figure}

Five main regions of different dynamic solutions are shown in Fig. \ref{figure2}. At low pressure values 
(region I), the system presents a stable fixed point. Reaching region II, the fixed point becomes unstable 
and there appears an attracting limit cycle. At the interface between regions I and II, three bifurcations occur 
in a narrow range of subglottal pressure (Fig. \ref{figure3}, left panel), all along the $Q$ axis.
The right panel of Fig. \ref{figure3} shows the oscillation amplitude of $x_2$. At point \textbf{A},
oscillations are born in a supercritical Hopf bifurcation. The amplitude grows continuously for increasing 
$P_s$ until point \textbf{B}, where it jumps to the upper branch. 
If the pressure is then decreased, the oscillations persist even for lower pressure values than the onset 
in \textbf{A}. When point \textbf{C} is reached, the oscillations suddenly stop and the system returns 
to the rest position. This onset-offset oscillation hysteresis was already reported experimentally in 
\cite{15_b}.

The branch \textbf{AB} depends on the viscosity. Decreasing $\mu$, points \textbf{A} and 
\textbf{B} approach to each other until they collide at $\mu=0$, recovering the result reported in \cite{7,14,9}, 
where the oscillations occur as the combination of a subcritical Hopf bifurcation and a cyclic fold bifurcation.

On the other hand, the branch \textbf{BC} depends on the separation point of the jet formation. In particular, 
for increasing $k_s$, the folds become stiffer and the separation point moves upwards toward the output 
of the glottis. From a dynamical point of view, points \textbf{C} and \textbf{B} approach to each other until 
they collapse. In this case, the oscillations are born at a supercritical Hopf bifurcation and the system presents 
no hysteresis, as in the standard flapping model \cite{19}. 

%\vspace{.2cm}

Regions II and III of Fig. \ref{figure2} are separated by a saddle-repulsor bifurcation. Although this bifurcation 
does not represent a qualitative dynamical change for the oscillating folds, 
its effects are relevant when the complete mechanism of voiced sound production is considered.
Voiced sounds are generated as the airflow disturbance produced by the oscillation of the vocal folds is
injected into the series of cavities extending from the laryngeal exit to the mouth, a non-uniform tube
known as the vocal tract.  
The disturbance travels back and forth along the vocal tract, that acts as a filter for the original signal,
enhancing the frequencies of the source that fall near the vocal tract resonances.
Voiced sounds are in fact perceived and classified according to these resonances, as in the case of vowels
\cite{16}. Consequently, one central aspect in the generation of voiced sounds is the production of a spectrally
rich signal at the sound source level.

\begin{figure}[!ht]
\begin{center}
\includegraphics[width=0.4\textwidth]{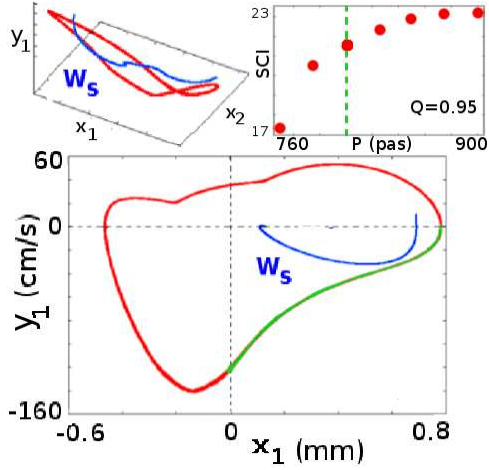}
\end{center}
\caption{A projection of the limit cycle for $x_1$ and the stable manifold of the saddle point, for parameters consistent
with normal phonatory conditions, $(Q,P_s)=(1,850)$ (region III). Left inset: projection in the 3-dimensional space ($y_1$, $x_1$,
$x_2$). Right inset: Spectral content index of $x_1(t)$ as a function of $P_s$ for a fixed value of $Q=0.95$. In green, the value  at which the 
saddle-repulsor bifurcation takes place.}
\label{figure4}
\end{figure}

Interestingly, normal phonation occurs in the region near the appearance of the saddle-repulsor bifurcation.
Although this bifurcation does not alter the dynamical regime of the system or its time scales, we have observed that 
part of the limit cycle approaches the stable manifold of the new fixed point (as displayed in Fig. \ref{figure4}), 
therefore changing its shape.
This deformation is not restricted to the appearance of the new fixed point but rather occurs in a coarse region 
around the boundary between II and III, as the flux changes smoothly in a vicinity of the bifurcation. 
In order to illustrate this effect, we use the spectral content index SCI \cite{gabo}, an indicator of the spectral 
richness of a signal: $SCI= \sum_k A_k f_k /(\sum_k A_k f_0)$, where $A_k$ is the Fourier amplitude of the 
frequency $f_k$ and $f_0$ is the fundamental frequency. 
As the pressure is increased, the SCI of $x_1(t)$ increases (upper right panel of Fig. \ref{figure4}), observing a 
boost in the vicinity of the saddle-repulsor bifurcation that stabilizes after the saddle point is generated.

Thus, the appearance of this bifurcation near the region of normal phonation could indicate a possible mechanism 
to further enhance the spectral richness of the sound source, on which the production of voiced sounds ultimately relies.

%\vspace{0.2cm}

In the boundary between regions III and IV, one of the unstable points created in the saddle-repulsor bifurcation undergoes
a subcritical Hopf bifurcation, changing stability as an unstable limit cycle is created \cite{18}. Finally, entering region V,
the stable and the unstable cycles collide and disappear in a fold of cycles  where no oscillatory regimes exist.

In Fig. 2, we also display a color map that quantifies the difference between the solutions of the model and the flapping 
approximation. The flapping model is a two dimensional model that, instead of two masses per fold, assumes a 
wave propagating along a linear profile of the folds, i.e., the displacement of the upper edge of the folds is delayed $2\tau$ with respect 
to the lower. The cross sectional areas at glottal entry and exit ($a_1$ and $a_2$) are approximated, in terms of the position 
of the midpoint of the folds, by

\begin{align}
\label{areas}
\left\{
\begin{array}{rl}
a_1 = 2 l_g (x_{0} + x + \tau \dot{x})	  \\
a_2 = 2 l_g (x_{0} + x - \tau \dot{x})	 \
\end{array} \right., 
\end{align}

\noindent where $x$ is the midpoint displacement from equilibrium $x_0$, and $\tau$ is the time that the surface wave 
takes to travel half the way from bottom to top. Equation (\ref{areas}) can be rewritten as $(x_1-x_2)=\tau (y_1+y_2)$. 
We use this expression to quantify the difference between the oscillations obtained with the  two-mass model solutions 
and the ones generated with the flapping approximation, computing the linear correlation coefficient between $(x_1-x_2)$ 
and $(y_1+y_2)$. As expected, the correlation coefficient $R$ decreases for increasing $P_s$ or decreasing $Q$. In the 
region near normal phonation, the approximation is still relatively good, with $R\sim 0.8$. As expected, the approximation 
is better for increasing $x_0$, since the effect of colliding folds is not included in the flapping model. 

\subsection{Isofrequency curves}
\label{isof}

One basic perceptual property of the voice is the pitch, identified with the fundamental frequency $f_0$ of the vocal folds
oscillation. The production of different pitch contours is central to language, as they affect the semantic content of
speech, carrying accent and intonation information. Although experimental data on pitch control is scarce, it was reported 
that it is actively controlled by the laryngeal muscles and the subglottal pressure. In particular, when the vocalis or 
interarytenoid muscle activity is inactive, a raise of the subglottal pressure produces an upraising of the pitch 
\cite{22}.

\begin{figure}[!ht]
\begin{center}
\includegraphics[width=0.47\textwidth]{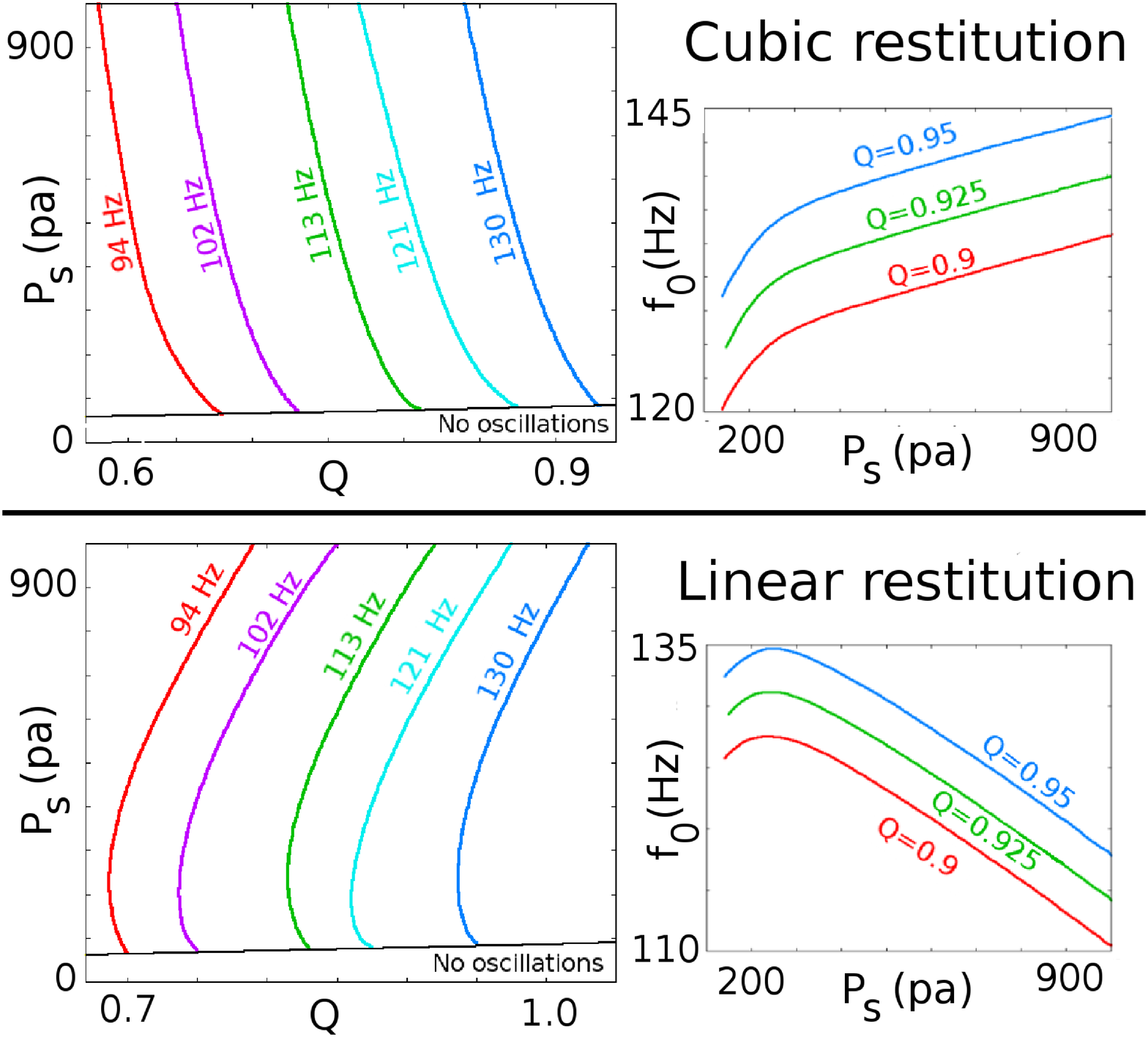}
\end{center}
\caption{Relationship between pitch and restitution forces. Left panels: isofrequency curves in the plane ($Q$,$P_s$). 
Right panels: Curves $f_0(P_s)$ for $Q$=0.9, $Q$=0.925 and $Q$=0.95. In  the upper panels, we
used the model with the cubic nonlinear restitution of Eq. (\ref{restitution}). In the lower panels, we show the curves
obtained with a linear restitution, $K_i(x_i)=k_i x_i+ \Theta(\frac{x_i+x_0}{x_0}) 3 k_i (x_i + x_0)$.}
\label{figure5}
\end{figure}

Compatible with these experimental results, we performed a theoretical analysis using $P_s$ as a single control parameter
for pitch. 
In the upper panels of Fig. \ref{figure5}, we show isofrequency curves in the range of normal speech for our model
of Eqs. (\ref{motion}) to (\ref{f2}). Following the ideas developed in \cite{Am08} for the avian case, we compare the 
behavior of the fundamental frequency with respect to pressure $P_s$ in the two most usual cases presented in the 
literature: the cubic \cite{1,12} and the linear \cite{14,9} restitutions. 
In the lower panels of Fig. \ref{figure5}, we show the isofrequency curves that result from replacing the cubic 
restitution by a linear restitution $K_i(x_i)=k_i x_i+ \Theta(\frac{x_i+x_0}{x_0}) 3 k_i (x_i + x_0)$.

Although the curves $f_0(P_s)$ are not affected by the type of restitution at the very beginning of oscillations, the 
changes become evident for higher values of $P_s$, with positive slopes for the cubic case and negative for the linear 
case. This result suggests that a nonlinear cubic restitution force is a good model for the elastic properties 
of the oscillating tissue. 

\section{Conclusions}
\label{conclusions}

In this paper, we have analyzed a complete two-mass model of the vocal folds integrating collisions, nonlinear restitution and 
dissipative forces for the tissue and jets and viscous losses of the air-stream. In a framework of growing interest for 
detailed modeling of voice production, the aspects studied here contribute to understanding  the role of the different 
physical terms in different dynamical behaviors. 

We calculated the bifurcation diagram, focusing in two regimes: the oscillation 
onset and normal phonation. Near the parameters of normal phonation, a saddle repulsor bifurcation takes place that modifies 
the shape of the limit cycle, contributing to the spectral richness of the glottal flow, which is central to the production
of voiced sounds. With respect to the oscillation onset, we showed how jets and viscous losses intervene in the hysteresis 
phenomenon.
    
Many different models for the restitution properties of the tissue  have been used across the literature, including linear and 
cubic functional forms. Yet, its specific role was not reported. Here we showed that the experimental relationship between 
subglottal pressure and pitch is fulfilled by a cubic term.

\begin{acknowledgements}
This work was partially funded by UBA and CONICET.
\end{acknowledgements}


\begin{thebibliography}{50} %This parameter should be a number with as many digits as the number of references

% Journal Articles: Comma separated Authors' names (initial, with no periods, and family name), Full title of the article (italic), Abbreviated journal title, volume (bold), first page (or article number), (year).
\bibitem{1} K Ishizaka, J L Flanagan, \textit{Synthesis of voiced sounds from a two-mass model of the vocal cords}, Bell Syst. Tech. J. \textbf{51}, 1233 (1972).

\bibitem{6} I R Titze, \textit{The physics of small‐amplitude oscillation of the vocal folds}, J. Acoust. Soc. Am. \textbf{83}, 1536 (1988).

\bibitem{7} M A Trevisan, M C Eguia, G Mindlin, \textit{Nonlinear aspects of analysis and synthesis of speech time series data},  Phys. Rev. E \textbf{63}, 026216 (2001).

\bibitem{2} Y S Perl, E M Arneodo, A Amador, F Goller,  G B Mindlin, \textit{Reconstruction of physiological instructions from Zebra finch song},  Phys. Rev. E \textbf{84}, 051909 (2011).

\bibitem{3} E M Arneodo, Y S Perl, F Goller, G B Mindlin, \textit{Prosthetic avian vocal organ controlled by a freely behaving bird based on a low dimensional model of the biomechanical periphery},  PLoS Comput. Biol. \textbf{8},  e1002546 (2012). 
%\bibitem{2} T. Shipp, R.E. McGlone, \textit{Laryngeal Dynamics Associated with Voice Frequency Change}, Journal of Speech and Hearing Research \textbf{14}, 761, (1971).
%\bibitem{3} T. Baer, \textit{Reflex activation of laryngeal muscles by sudden induced subglottal pressure changes}, J. Acoust. Soc. Am. \textbf{65}, 1271, (1979).
%\bibitem{4} H. Teager, \textit{Some observations on oral air flow during phonation}, Acoustics, Speech and Signal Processing \textbf{28}, 599, (1980).
%\bibitem{5} I.R. Titze, H. Liang, \textit{Comparison of Fo extraction methods for high-precision voice perturbation measurements}, Journal of Speech and Hearing Research \textbf{36}, 1120, (1993).

\bibitem{11}  B H Story, I R Titze \textit{Voice simulation with a body‐cover model of the vocal folds}, J. Acoust. Soc. Am. \textbf{97}, 1249 (1995).

\bibitem{12} J C Lucero, L Koening \textit{Simulations of temporal patterns of oral airflow in men and women using a two-mass model of the vocal folds under dynamic control}, J. Acoust. Soc. Am. \textbf{117}, 1362 (2005).

\bibitem{17}  X Pelorson, X Vescovi, C Castelli, E Hirschberg, A Wijnands, A P J Bailliet, H M A Hirschberg, \textit{Description of the flow through in-vitro models of the glottis during phonation. Application to voiced sounds synthesis},  Acta  Acust. \textbf{82}, 358 (1996).

\bibitem{13}  M E Smith, G S Berke, B R Gerratt, \textit{Laryngeal paralyses: Theoretical considerations and effects on laryngeal vibration},  J. Speech Hear. Res. \textbf{35}, 545 (1992).

\bibitem{14} I Steinecke, H Herzel \textit{Bifurcations in an asymmetric vocal‐fold model}, J. Acoust. Soc. Am. \textbf{97}, 1874 (1995).

\bibitem{15} N J C Lous, G C J Hofmans, R N J Veldhuis, A Hirschberg, \textit{A symmetrical two-mass vocal-fold model coupled to vocal tract and trachea, with application to prosthesis design},  Acta Acust. United Ac. \textbf{84}, 1135 (1998).

\bibitem{15_b} T Baer, \textit{Vocal fold physiology}͑, University of Tokyo Press, Tokyo, (1981).

\bibitem{8} T Ikeda, Y Matsuzak, T Aomatsu, \textit{A numerical analysis of phonation using a two-dimensional flexible channel model of the vocal folds},  J. Biomech. Eng. \textbf{123}, 571 (2001).

\bibitem{9} J C Lucero, \textit{Dynamics of the two-mass model of the vocal folds: Equilibria, bifurcations, and oscillation region}, J. Acoust. Soc. Am. \textbf{94}, 3104 (1993).

\raggedbottom 
\pagebreak

\bibitem{10}  X Pelorson, A Hirschberg, R R van Hassel, A P J Wijnands, Y Auregan, \textit{Theoretical and experimental study of quasisteady‐flow separation within the glottis during phonation. Application to a modified two‐mass model}, J. Acoust. Soc. Am. \textbf{96}, 3416 (1994).

%\bibitem{16} T. Baer, T., \textit{Observation of vocal fold vibration: Measurement of excised larynges}, In: Vocal Fold Physiology, Eds. K.N. Stevens and M. Hirano, pag. 119, ͑University of Tokyo, Tokyo͒, (1981).

\bibitem{22}  T Baer, \textit{Reflex activation of laryngeal muscles by sudden induced subglottal pressure changes}, J. Acoust. Soc. Am. \textbf{65}, 1271 (1979).



% Books: Comma separated Authors' names (initial, with no periods, and family name), Full title of the book (italic), Editorial, City, (year).
\bibitem{19} J C Lucero, \textit{A theoretical study of the hysteresis phenomenon at vocal fold oscillation onset-offset}, J. Acoust. Soc. Am. \textbf{105}, 423 (1999). 

\bibitem{16} I Titze, \textit{Principles of voice production}, Prentice Hall, (1994).

\bibitem{18} J Guckenheimer, P Holmes, \textit{Nonlinear oscillations, dynamical systems and bifurcations of vector fields}, Springer, (1983).

%\bibitem{21} T. Shipp, R.E. McGlone, \textit{Laryngeal Dynamics Associated with Voice Frequency Change}, Journal of Speech and Hearing Research \textbf{14}, 761, (1971).

\bibitem{20} E Doedel, \textit{AUTO: Software for continuation and bifurcation problems in ordinary differential equations}, AUTO User Manual, (1986).

\bibitem{gabo} J Sitt, A Amador, F Goller,  G B Mindin, \textit{Dynamical origin of spectrally rich vocalizations in birdsong}, Phys. Rev. E \textbf{78}, 011905 (2008).  

\bibitem{Am08} A Amador, F Goller, G B Mindlin, \textit{Frequency modulation during song in a suboscine does not require vocal muscles},  J. Neurophysiol. \textbf{99}, 2383 (2008).

\end{thebibliography}
\end{document}